\PassOptionsToPackage{unicode}{hyperref}
\PassOptionsToPackage{hyphens}{url}
\documentclass[
]{amsart}
\usepackage{lmodern}
\usepackage{amssymb,amsmath}
\usepackage{ifxetex,ifluatex}
\ifnum 0\ifxetex 1\fi\ifluatex 1\fi=0 
  \usepackage[T1]{fontenc}
  \usepackage[utf8]{inputenc}
  \usepackage{textcomp} 
\else 
  \usepackage{unicode-math}
  \defaultfontfeatures{Scale=MatchLowercase}
  \defaultfontfeatures[\rmfamily]{Ligatures=TeX,Scale=1}
\fi
\IfFileExists{upquote.sty}{\usepackage{upquote}}{}
\IfFileExists{microtype.sty}{
  \usepackage[]{microtype}
  \UseMicrotypeSet[protrusion]{basicmath} 
}{}

\usepackage{xcolor}
\IfFileExists{xurl.sty}{\usepackage{xurl}}{} 
\IfFileExists{bookmark.sty}{\usepackage{bookmark}}{\usepackage{hyperref}}
\hypersetup{
  hidelinks,
  pdfcreator={LaTeX via pandoc}}
\usepackage{longtable,booktabs}
\usepackage{etoolbox}
\makeatletter
\patchcmd\longtable{\par}{\if@noskipsec\mbox{}\fi\par}{}{}
\makeatother
\IfFileExists{footnotehyper.sty}{\usepackage{footnotehyper}}{\usepackage{footnote}}
\makesavenoteenv{longtable}
\usepackage{graphicx,grffile}
\makeatletter
\def\maxwidth{\ifdim\Gin@nat@width>\linewidth\linewidth\else\Gin@nat@width\fi}
\def\maxheight{\ifdim\Gin@nat@height>\textheight\textheight\else\Gin@nat@height\fi}
\makeatother
\setkeys{Gin}{width=\maxwidth,height=\maxheight,keepaspectratio}
\makeatletter
\def\fps@figure{htbp}
\makeatother
\ifxetex
  \usepackage{bidi}
\fi
\ifnum 0\ifxetex 1\fi\ifluatex 1\fi=0 
  \newcommand{\LR}[1]{\beginL #1\endL}

\fi

\title{Testing for trends in online experiments}

\author{Chris Haulk}
\email{haulk@google.com}
\author{Lee Richardson}
\email{leerich@google.com}
\author{Jacopo Soriano}
\email{jacoposoriano@google.com}
\address{Google}
\date{}

\usepackage[numbers]{natbib}
\usepackage{float}
\floatstyle{ruled}
\newfloat{algorithm}{htbp}{loa}
\floatname{algorithm}{Algorithm}
\begin{document}

\begin{abstract}
Time-varying treatment effects can be estimated with linear regression
paired with jackknife confidence intervals. In experiments at a large
online video platform, the test based on this estimator is not much less
powerful than modern tests for monotonic trends. If you suspect trends
in your online experiments, consider using this estimator.
\end{abstract}
\maketitle

\hypertarget{introduction}{%
\section{\texorpdfstring{\LR{Introduction}}{Introduction}}\label{introduction}}
\LR{Time-dependent treatment effects -\/- i.e. \emph{trends} -\/- are
common in online experiments, and important to notice. Indeed, trends in
treatment effects are strong proxies or surrogates in the sense of
\cite{tripuraneni, zito, sigerson} for long-term outcomes in general, and upward
trends are considered the hallmark of effective treatments at YouTube. A
common method for assessing trends is to ``eyeball the time series'',
i.e. to visually check whether the plot of treatment effects vs time has
an ``up-and-to-the-right'' shape. This was standard practice in our
company for many years and seems to be a supported option in several
online testing platforms: for example, Statsig provides a time series
view to assist with this task \cite{statsig}. However, ``eyeballing a
time series'' is not very statistically rigorous, and it does not scale
to the large number of metrics we monitor in our online experiments.}

\LR{``Cookie/cookie-day'' experiments, introduced in \cite{hohnhold2015focusing}, are
an effective experiment design for assessing} \LR{trends. The idea is to
treat some units continuously, and others only at one instant, allowing
the estimation of both the cumulative treatment effect and the
instantaneous effect. The difference between these is an estimate of the
``time-dependent effect'' or trend. The cookie/cookie-day experiment
design allows the estimation of a wide variety of possible trends, which
gives it great appeal.}

\LR{The authors of \cite{hohnhold2015focusing} show empirically that time-dependent
treatment effects can take many interesting nonlinear patterns in the
online advertising context. But in our experience working on search and
recommendation systems at YouTube, most trends are smooth, monotonic and
essentially linear over moderate time horizons. When this is so, they
can be detected from a ``usual'' experiment design in which all units
are treated continuously, simply by regressing treatment effects
estimates on the time elapsed since the start of the experiment. This
estimator is easy to implement, integrates well with covariate
adjustment techniques \cite{soriano2017pre, deng2013improving, fisher1932statistical}, and requires no
change to usual experiment design. Moreover, the regression coefficient
it provides -\/- the estimated ``change in treatment effect per unit
time'' -\/- is a simple, interpretable summary that puts ``trend'' on
the same footing as ``treatment effect'', which can be useful for
response-surface-methodology applications \cite{haulk} even when a
straight-line trend model is not exactly correct. This regression-based
trend estimator can perform well even when trend effects are not exactly
linear. We demonstrate this below in a simulation comparing the
performance of this estimator against a more modern test for monotonic
trends.}

\LR{Treatment effect estimates are often correlated across time,
invalidating the ``naive'' standard errors and confidence intervals for
regression-based trend estimates. Classic references on this include
\cite{yule1926why} and \cite{cochrane1949application}. Possible fixes include estimating
correlation structure and using weighted least squares, but for
simplicity we prefer to use the jackknife estimate of variance,
explained below and in \cite{lipsitz1990using}.}

\LR{Internally at YouTube, we call our estimate of trend the jackknife
linear model. It is implemented in Google's centralized experimentation
platform, and is used over 10,000 times per day. It is a practical and
widely-used tool for analyzing trends in online experiments, and if your
experiment platform does not support trend estimation, you might
consider giving this method a try.}

\vspace{0.15in}

\noindent \includegraphics[width=\textwidth]{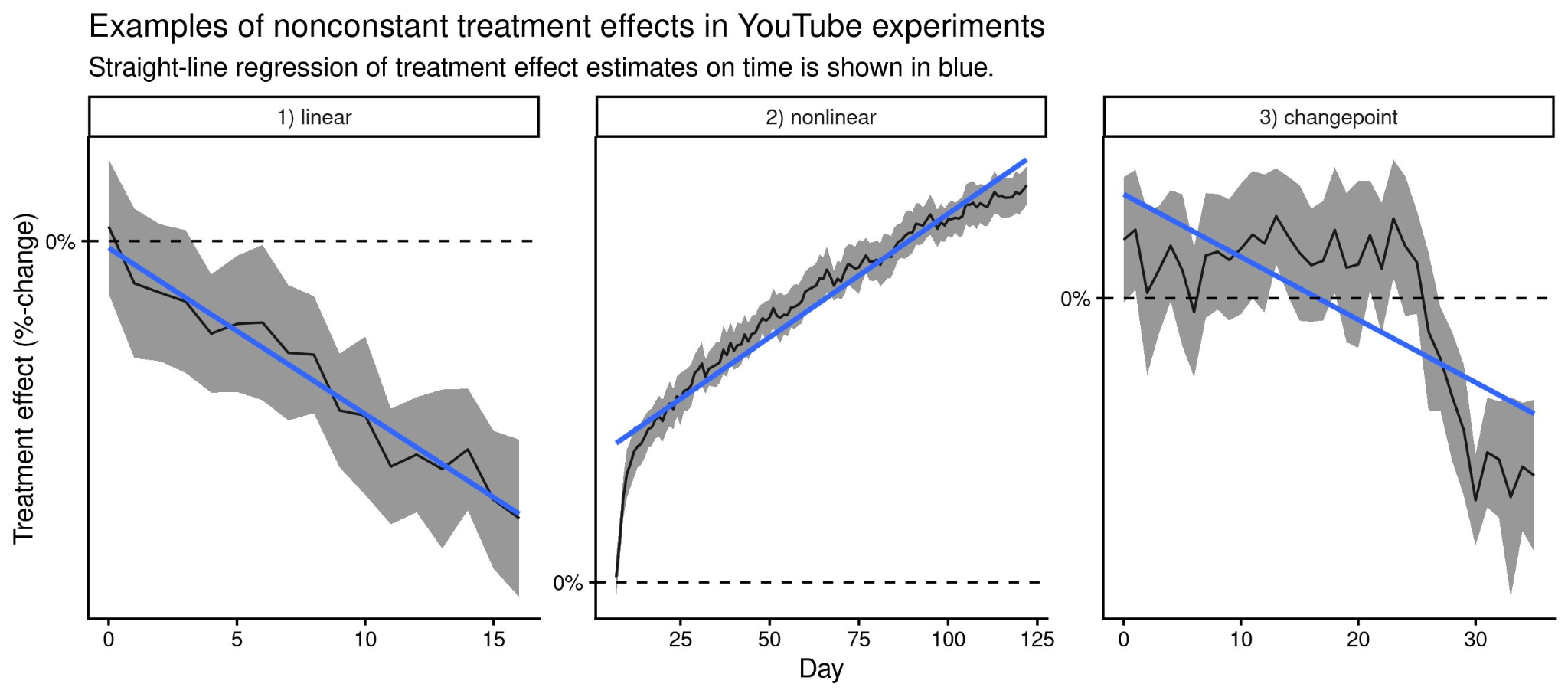}

\vspace{0.15in}

\hypertarget{trend-estimation-via-jackknife-linear-model}{%
\section{\texorpdfstring{\LR{Trend estimation via jackknife linear
model}}{Trend estimation via jackknife linear model}}\label{trend-estimation-via-jackknife-linear-model}}
\LR{As explained in \cite{tang2010overlapping}, YouTube experiment data is aggregated
and anonymized into sufficient statistics before it is further analyzed.
The unit of analysis, a ``cookie bucket'', represents a group of many
users, usually hundreds of thousands or millions, and cookie bucket
measurements represent the sum of a simple quantity over all users in
the group on a particular day, like ``number of YouTube videos viewed''
or ``total time spent watching YouTube videos''. We call these outcome
measures ``metrics''. For the following discussion we suppose users are
stably assigned to cookie buckets and followed over time in an
experiment with two arms.}
\LR{Let $y_{a, k, d}$ represent the metric value associated with the
$k$-th cookie-bucket assigned to arm $a$ on day $d$ of the experiment, $a=0, 1$,
$k=0\ldots K-1$, $d=0\ldots D-1$. And let $x_{a, k}$ represent a covariate
associated with the $k$-th cookie bucket in arm $a$, and let $x$ =
$(2K)^{-1} \sum_{k,a} x_{a,k}$ be its average. We write}

\begin{quote}
\LR{$y_{a, k, d}$ = $\mu_{0, d}$ + $\mu_{1, d}$ $\times$ 1(a=1) + $\mu_{2, a,
d}$ $\times$ ($x_{a, k}$ - $x$) + $\eta_{a, k, d}$,}
\end{quote}
\LR{and estimate parameters $\mu$ by least squares, providing an
ANCOVA-style estimate of the per-day baseline rate $\mu_{0, d}$ and the
additive effect of treatment on that day $\mu_{1, d}$. The ratio}
\begin{quote}
\LR{$\hat{\Delta}_{d}$ = $\hat{\mu}_{1, d}$ /
$\hat{\mu}_{0, d}$}
\end{quote}
\LR{is the PrePost estimate of the treatment effect in relative change
units on day $d$ of the experiment \cite{soriano2017pre}. Time series of treatment
effects in our online experiments often seem aptly described by a
straight-line model}
\begin{quote}
\LR{$\hat{\Delta}_{d} = b_{0} + b_{1} \times d + \epsilon_{d}$.}
\end{quote}

\LR{We call $b_{0}$ and $b_{1}$ in this equation \emph{lift} and \emph{trend},
respectively, and we estimate them by ordinary least-squares. Letting $\hat{\Delta}$
denote the vector $[\hat{\Delta}_{d}, d=0, \ldots, D-1]$ and $F$ the matrix whose $d$-th
row is $[1, d]$, the ordinary least squares (OLS) estimate of lift and
trend is}
\begin{quote}
\LR{$\hat{b} = [\hat{b}_0, \hat{b}_1]^T = (F^TF)^{-1} F^T\hat{\Delta}$,}
\end{quote}
\LR{and the OLS estimate of their variance is}
\begin{quote}
\LR{$\widehat{\text{Cov}}([\hat{b}_0, \hat{b}_1]) = \hat{\sigma}^2
(F^TF)^{-1}$}
\end{quote}
\LR{where $\hat{\sigma}^2 = || \hat{\Delta} - F \hat{b} ||^2 / (D-2)$. This OLS variance
estimate is badly biased on YouTube data, as discussed below, because
the errors $\epsilon_{d}$ in the straight-line trend model are
correlated. However, different cookie-buckets comprise different users
assigned completely at random, so the vectors $y_{k}$ = $[y_{a, k, d} : a=0, 1; d=0\ldots D-1]$ are approximately independent and identically
distributed. Thus we rely on the} \LR{jackknife estimate of variance
\cite{efron1981jackknife}: essentially we resample cookie-buckets to estimate
uncertainty. At YouTube we refer to this estimator of lift, trend and
error \emph{the jackknife linear model}.}
\begin{algorithm}[H]
\caption{Jackknife Linear Model Trend Estimation}
\begin{enumerate}
    \item Let $\hat{\Delta}^{(-k)}$ denote the vector of per-day treatment effect estimates based on all data except the data from the $k$-th cookie bucket
    \item $\hat{b}^{(-k)} = (F^TF)^{-1} F^T\hat{\Delta}^{(-k)}$, the estimated lift and trend based on all data except the data from the $k$-th cookie bucket
    \item $\widehat{\text{Cov}}([\hat{b}_0, \hat{b}_1]) = \frac{K-1}{K} \sum_k (\hat{b}^{(-k)} - \hat{b}) (\hat{b}^{(-k)} - \hat{b})^T$
\end{enumerate}
\end{algorithm}
\hypertarget{ols-variance-estimates-of-trend-are-biased-on-youtube-data}{%
\section{\texorpdfstring{\LR{OLS variance estimates of trend are biased
on YouTube
Data}}{OLS variance estimates of trend are biased on YouTube Data}}\label{ols-variance-estimates-of-trend-are-biased-on-youtube-data}}
\LR{As mentioned above, treatment effect estimates are often correlated
across time, invalidating the ``naive'' OLS standard errors and
confidence intervals for $\hat{b}_{1}$. We can demonstrate this using A/A
experiments.} \LR{The figure below at left shows a scatterplot of [$\hat{b}_{0}$, $\hat{b}_{1}$] for 1000 A/A experiments. Because no treatment is applied, [$b_{0}$, $b_{1}$] = (0, 0) in each experiment, and because experiments are
identically configured -\/- each ran for 14 days, for example -\/- the
variance-covariance matrix of [$\hat{b}_{0}$, $\hat{b}_{1}$] is the same in each
experiment. The ellipses shown in the plot were produced as follows: in
each A/A experiment we estimated the variance-covariance matrix of
[$\hat{b}_{0}$, $\hat{b}_{1}$] with OLS and again with the jackknife. Averaging yields
$\Sigma_{\text{OLS}}$ and $\Sigma_{\text{Jackknife}}$ , which are the
average variance estimates across A/A experiments. We also computed
$\Sigma_{\text{Empirical}}$ , the empirical variance-covariance for the
1000 lift and trend estimates [$\hat{b}_{0}$, $\hat{b}_{1}$]. The ellipses in the plots
are points $b$ for which $b^T \Sigma^{-1} b = 5.991$, a value
chosen because $b \sim \text{Normal}(0, \Sigma)$ implies $\text{Prob}(b^T \Sigma^{-1} b \le 5.991) = 0.95$. The figure below at
right was produced in the same way, except using PrePost covariate
adjustments to the time series ($y_{d}$) of daily treatment-effect
estimates.}

\vspace{0.15in}

\noindent \includegraphics[width=\textwidth]{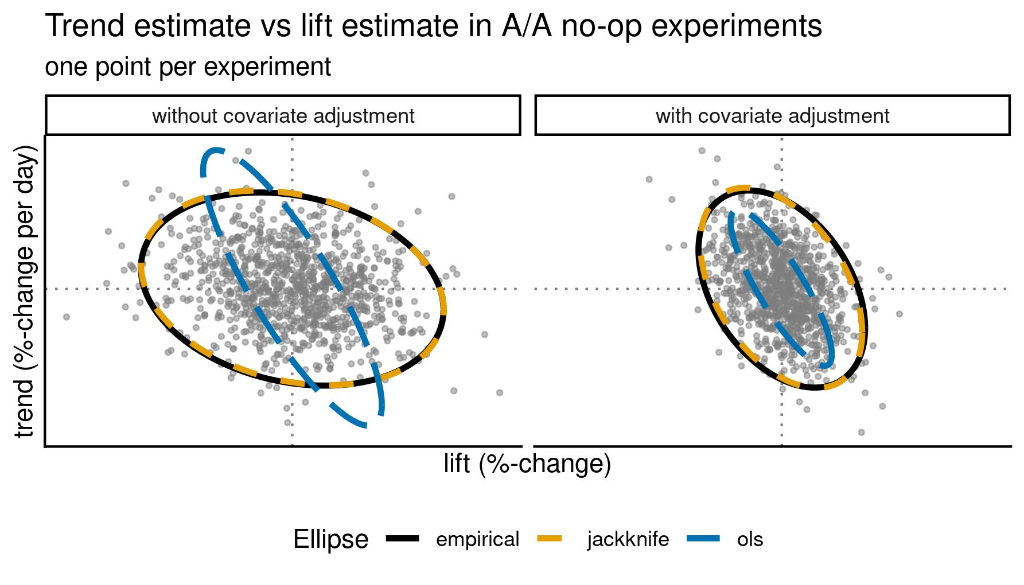}

\vspace{0.15in}

\LR{The agreement between the \emph{jackknife} and \emph{empirical}
ellipses demonstrates that the jackknife estimate of variance is
unbiased, as expected from \cite{efron1981jackknife}. Likewise, the difference
between the \emph{ols} and empirical ellipses demonstrates that the OLS
variance estimates are biased. Comparing widths and heights of the two
``empirical'' ellipses reveals that, in YouTube experiment data,
covariate adjustment sharpens estimates of the day-0 lift effect, but
does not affect precision of estimates of trend.} \LR{}
\LR{Bias of OLS variance estimates is quantified in the table below, in
which the ``target'' column contains empirical estimates of standard
deviations from 1000 A/A tests, or 0.95 in the case that the estimand is
observed coverage of a t-statistic-based confidence region with 95\%
nominal coverage. Four such regions are evaluated:}
\begin{itemize}
\item
  \begin{quote}
\LR{confidence interval for lift $b_{0}$}
\end{quote}
\item
  \begin{quote}
\LR{confidence interval for trend $b_{1}$}
\end{quote}
\item
  \begin{quote}
\LR{a joint confidence region for pair [$b_{0}$, $b_{1}$]}
\end{quote}
\item
  \begin{quote}
\LR{a confidence interval for the 28d \emph{extrapolation,} $b_{0}$ + $b_{1}$ $\times$ 27.}
\end{quote}
\end{itemize}
\LR{The table quantifies bias in the OLS variance estimates and the
resulting coverage errors. It also shows that covariate adjustment
sharpens estimates of lift (cutting lift\_sd by about 50\%).}

\begin{longtable}[]{@{}lllllll@{}}
\toprule
\shortstack[l]{cov\\adj} & \shortstack[l]{estimand} & \shortstack[l]{target} &
\shortstack[l]{ols\\est} & \shortstack[l]{ols\\rel\\error\\(\%)} &
\shortstack[l]{jackknife\\est} &
\shortstack[l]{jackknife\\rel\\error\\(\%)}\tabularnewline
\midrule
\endhead
\LR{FALSE} & \LR{lift se} & \LR{0.0082} & \LR{0.0048} & \LR{-41} & \LR{0.0082} & \LR{0.055}\tabularnewline
\LR{FALSE} & \LR{trend se} & \LR{0.00044} & \LR{0.00063} & \LR{43} & \LR{0.00045} & \LR{0.69}\tabularnewline
\LR{FALSE} & \LR{extrap 28 se} & \LR{0.013} & \LR{0.014} & \LR{3.9} & \LR{0.013} & \LR{-0.93}\tabularnewline
\LR{FALSE} & \LR{lift coverage} & \LR{0.95} & \LR{0.76} & \LR{-20} & \LR{0.96} & \LR{0.53}\tabularnewline
\LR{FALSE} & \LR{trend coverage} & \LR{0.95} & \LR{0.99} & \LR{4.1} & \LR{0.94} & \LR{-0.63}\tabularnewline
\LR{FALSE} & \LR{extrap 28 coverage} & \LR{0.95} & \LR{0.95} & \LR{-0.21} & \LR{0.95} & \LR{-0.11}\tabularnewline
\LR{FALSE} & \LR{joint coverage} & \LR{0.95} & \LR{0.55} & \LR{-42} & \LR{0.95} & \LR{0.0}\tabularnewline
\LR{TRUE} & \LR{lift se} & \LR{0.0045} & \LR{0.0027} & \LR{-39} & \LR{0.0044} & \LR{-2.8}\tabularnewline
\LR{TRUE} & \LR{trend se} & \LR{0.00045} & \LR{0.00036} & \LR{-21} & \LR{0.00046} & \LR{1.6}\tabularnewline
\LR{TRUE} & \LR{extrap 28 se} & \LR{0.012} & \LR{0.0078} & \LR{-33} & \LR{0.012} & \LR{-0.60}\tabularnewline
\LR{TRUE} & \LR{lift coverage} & \LR{0.95} & \LR{0.80} & \LR{-16} & \LR{0.94} & \LR{-1.3}\tabularnewline
\LR{TRUE} & \LR{trend coverage} & \LR{0.95} & \LR{0.90} & \LR{-4.9} & \LR{0.96} & \LR{0.63}\tabularnewline
\LR{TRUE} & \LR{extrap 28 coverage} & \LR{0.95} & \LR{0.84} & \LR{-12} & \LR{0.95} & \LR{0.11}\tabularnewline
\LR{TRUE} & \LR{joint coverage} & \LR{0.95} & \LR{0.56} & \LR{-41} & \LR{0.94} & \LR{-0.84}\tabularnewline
\bottomrule
\end{longtable}

\vspace{0.15in}

\hypertarget{an-alternative-trend-estimator-and-jackknife-linear-model-performance-under-model-misspecification}{%
\section{\texorpdfstring{\LR{An alternative trend estimator, and
jackknife-linear-model performance under model
misspecification}}{An alternative trend estimator, and jackknife-linear-model performance under model misspecification}}\label{an-alternative-trend-estimator-and-jackknife-linear-model-performance-under-model-misspecification}}
\LR{Here we present a simulation study comparing the regression test for
trends to one specifically for monotonic trends. The generative model in
our simulation is based on our A/A experiments:} \LR{we take an A/A
experiment and inject a fake trend in treatment arm, of the form}
\begin{quote}
\LR{$\mu_{1,d}$ = s $\times$ $m_d$ $\times$ $(d/(D-1))^{(1/p)}$}
\end{quote}
\LR{for p=1, 2, 3. For p=1 the trend is linear, and p \textgreater{} 1
it is increasing and concave downward. We also try simulations for ``no
trend'' $\mu_{1,d}$ = 0, and ``changepoint''
$\mu_{1,d}$ = 1 if d\textgreater=7 else 0. Here
$m_d$ represents an estimate of the baseline rate $\mu_{0,
d}$ across all of our A/A tests and s is a scale parameter whose value
is chosen to make the power of our trend tests roughly 50\%.}
\LR{The figure below shows some samples from this generative model and the mean
function to illustrate the trend.}

\vspace{0.15in}

\noindent \includegraphics[width=\textwidth]{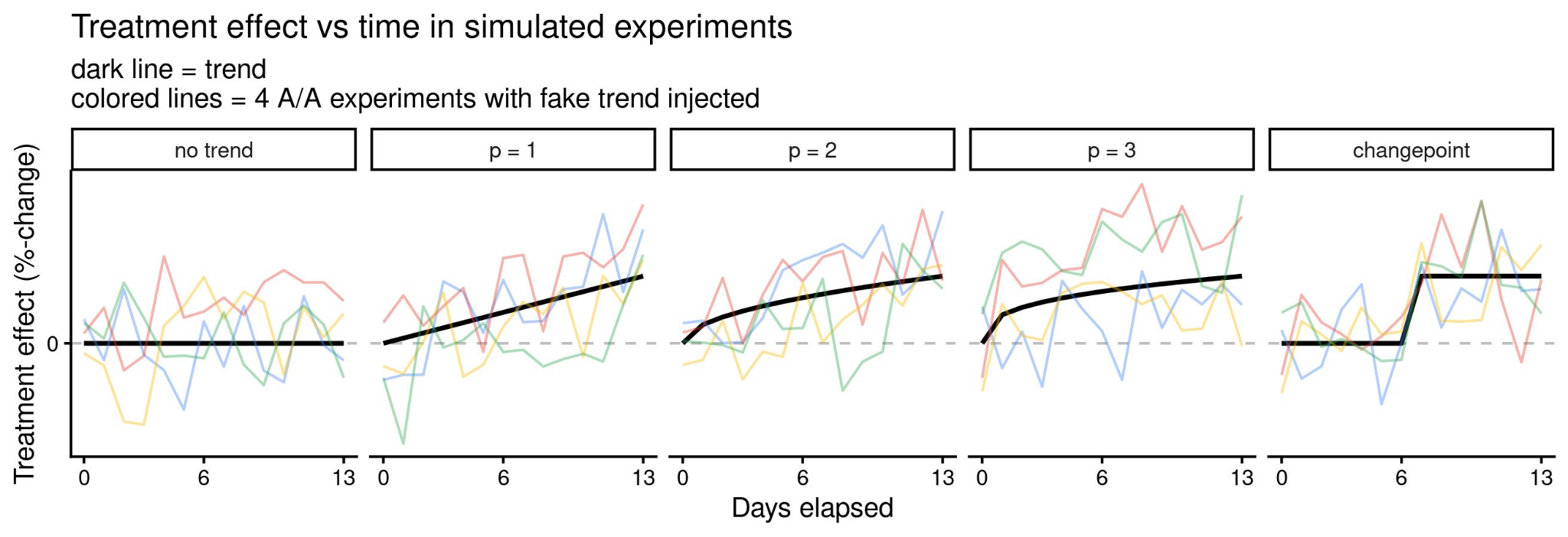}

\vspace{0.15in}

\LR{The competing test we evaluate in the simulation is based on a
constrained MLE, motivated by \cite{bartholomew1959test}, and it evaluates the
\emph{monotonic increasing} alternative against \emph{constant effect}
null hypothesis.}
\LR{To explain this test, let $r_{a, k, d}$ denote the residual after
regression adjustment for the covariate,}
\begin{quote}
\LR{$r_{a, k, d}$ = $y_{a, k, d}$ - $\hat{\mu}_{2, a,
d}$$\times$($x_{a, k}$ - $x$).}
\end{quote}
\LR{If covariates are unavailable or unused, setting $r_{a, k, d}$ =
$y_{a, k, d}$ is fine. Then define differences and per-day average
differences,}
\begin{quote}
\LR{$\delta_{k, d}$ = $r_{1, k, d}$ - $r_{0, k, d}$}\\
\LR{$\delta_{d} = K^{-1} \sum \delta_{k,d}$.}
\end{quote}
\LR{Finally define the estimates of the per-day mean under $H_{0}$,}

\begin{quote}
\LR{$\hat{m} = D^{-1} \sum \delta_{d}$}
\end{quote}
\LR{and per-day mean under $H_{1}$, ($\hat{m}_0$ \ldots,
$\hat{m}_{D-1}$), which may be computed by isotonic regression or
the Pool-Adjacent-Violators Algorithm. Then the test statistic T is proportional to the log likelihood ratio under the assumption of IID errors,}
\begin{quote}
\LR{$T = \sum_{d} (\delta_{d} - \hat{m})^2 - \sum_{d} (\delta_{d} - \hat{m}_d)^2$.}
\end{quote}

\LR{To find the critical value for the test, we use a cluster bootstrap.
Specifically, to make the b-th bootstrap sample, we choose K
cookie-buckets $\{i_1^b, \ldots, i_K^b\}$ with replacement from the set of
possible cookie buckets \{1, \ldots, K\}, and form a new data set
via residual scaling, $\delta_{k,d}^{b} = \hat{m} + \sqrt{\frac{K}{K-1}} (\delta_{i_k^b, d} - \delta_d)$, and compute T on this data set,
$t^b$ = T(\{$\delta_{k,d}^{b}$, k=1..,K,
d=0..D-1\}), and repeat this process 1000 times. We use the upper
1-alpha quantile of $\{t^1, \ldots, t^{1000}\}$ as our critical value, i.e. the rejection
value for our test. We call this Bartholomew's bootstrap test or simply the PAVA test after the isotonic
regression algorithm.}
\LR{The table below shows rejection rates for our ``injected trends''
simulation. Because the PAVA test for trends is one-sided, we use a
one-sided t-test in our jackknife linear model.}

\vspace{0.15in}

\begin{longtable}[]{@{}lll@{}}
\toprule
\LR{trend (p)} & \LR{jackknife linear model rejection rate} & \LR{PAVA
rejection rate}\tabularnewline
\midrule
\endhead
\LR{no trend} & \LR{0.049} & \LR{0.05}\tabularnewline
\LR{linear, p=1} & \LR{0.598} & \LR{0.581}\tabularnewline
\LR{non linear, p=2} & \LR{0.487} & \LR{0.498}\tabularnewline
\LR{very nonlinear, p=3} & \LR{0.412} & \LR{0.452}\tabularnewline
\LR{changepoint} & \LR{0.851} & \LR{0.859}\tabularnewline
\bottomrule
\end{longtable}

\vspace{0.15in}

\LR{Our simulations indicate that the jackknife linear
model is capable of detecting nonlinear trends and changepoints in our online experiments, and its
power is close to that of the PAVA test. The PAVA test is slightly more
powerful when the trend is non-linear, but significantly more
complicated to implement, and it does not yield a simple, interpretable
estimate of the magnitude of the trend, whereas the straight-line regression model does.}

\hypertarget{conclusion}{%
\section{\texorpdfstring{\LR{Conclusion}}{Conclusion}}\label{conclusion}}
\LR{Time-dependent treatment effects -\/- which is to say, trends -\/- are common and
important in online experiments as pointed out in \cite{hohnhold2015focusing}, but
the literature on trend detection in online experimentation seems
sparse. In experiments with monotonic trends, regression-based estimates
and tests may be worth considering because they do not require the
cookie-cookie-day experiment design of \cite{hohnhold2015focusing}, they are easy to
execute, they provide simple numeric change-per-unit-time trend
summaries, and, depending on the context in which they are applied, they
may not lose much power relative to modern alternatives
\cite{bartholomew1959test, kudo1963multivariate, barlow1972statistical}. And if the trends in your online experiments are more complicated than the nearly-straight-line trends that we usually see at YouTube, there are of course textbook descriptions of flexible polynomial and spline models that might be suitable \cite{fitzmaurice2012applied}.}
\bibliographystyle{plainnat}
\bibliography{references}
\end{document}